# MODELING OF MOBILE VEHICLE SKID IN TRACTION MOVEMENT MODE

***Dmytro Klets,*** *Ph.D., e-mail: d.m.klets@gmail.com, prof_777@mail.ru*
*Kharkiv National Automobile and Highway University*

**Summary.** In the paper, a mathematical model assembling a "driver - mobile vehicle - road environment" system and capable of simulating the process of mobile vehicles skid in traction movement mode is proposed. The usage of non-linear drift models allows the development of efficient algorithms for mobile vehicles dynamic stabilization systems.

## INTRODUCTION

The traction loss between the vehicle and the road and skid is a reason of significant number of road accidents nowadays [2, 3, 6, 7, 9, 10, 16].

The creation of reliable and secure mobile vehicle involves the development and simulation of relevant mathematical models during the initial design [13, 14, 18] and in the course of its operation. The usage of non-linear models of the mobile vehicle skidding allows to estimate the influence of "driver - mobile vehicle - road environment" (DMVRE) system parameters on its movement, and to develop efficient algorithms for dynamic stabilization systems.

## PREVIOUS WORKS AND PUBLICATIONS

The design of an adequate model of the mobile vehicle during skidding is impossible without taking into account the interaction of elastic wheel with the ground. Several publications like [1, 8, 11, 20] are devoted to modeling of the tire contact with a road.

The authors of [11] have considered the process of elastic wheel moving after a sudden decrease of friction coefficient and vehicle full slip. The resulting dependence of the wheel linear and angular velocities, as well as its slip on the time allowed to determine the time of the wheel axle linear velocity fall from the initial value to zero. The authors of [17] have noted that the common experience of externally-mounted tractor facilities operation concurs the fact that tractors are predisposed for intensive oscillations of vehicle body during its normal operation.

In the work [15] it is indicated that the effective mobile vehicle skid damping is possible by means of automatic devices usage with large processing speed compared to the driver. The mobile vehicle dynamic stability systems use the possibility of side-by-side separate braking the wheels of various boards of mobile vehicles to create a stabilizing moments.

In the work [19], the linear and non-linear models of the skidding process are developed, considering the case of a vehicle braking. To obtain a plot of mobile vehicle line speeds the method of determining its instantaneous center of velocity (rotation) is proposed. The authors of this work accepted the hypothesis that the projection of a point of velocity center on the longitudinal axis of all-wheel drive mobile vehicle is outside the base (in front of the front axis). Otherwise, the front axle must slide laterally toward the center of rotation *O*. With all the wheels locked when the mobile vehicle has the kinetic energy, the front axle will tends to keep its stable run.

However, the issues of mobile vehicle motion modeling during skid in the traction mode require some additional research.

## THE OBJECTIVE

The aim of the research is to determine the angular acceleration of the mobile vehicle in the plane of the road, depending on its geometrical properties, the road surface traction rate, the linear and angular velocities of the mobile vehicle, as well as the wheel slip angles.

To achieve this goal it is necessary to compile and examine the differential equations of mobile vehicle plane motion during drift in the traction mode.

## THE SKID MODELING

Fig. 1 is a diagram of the forces acting on the rear-drive mobile vehicle in the process of skidding when driving in traction movement mode.

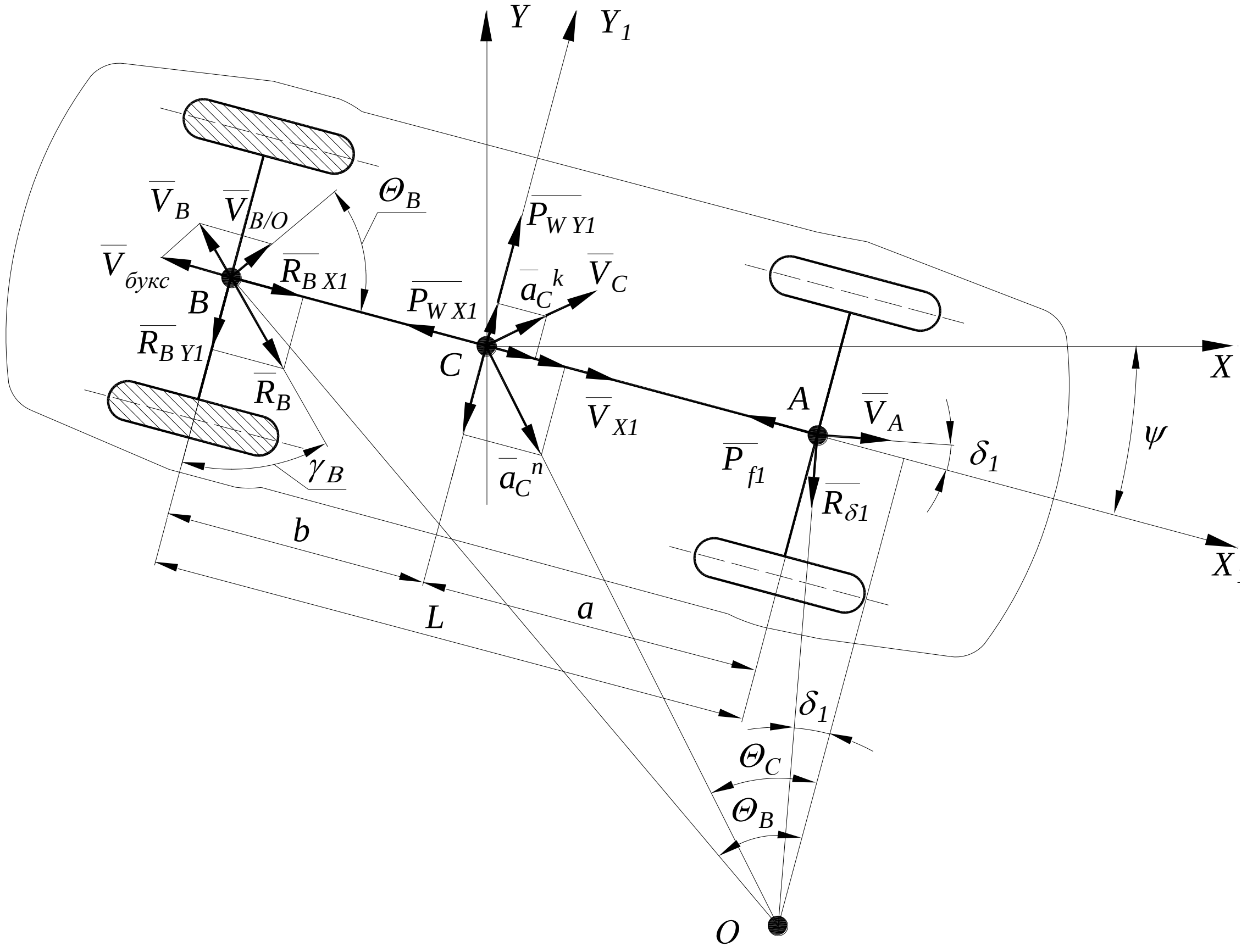

Fig. 1 – The diagram of the forces acting on the rear-drive mobile vehicle during skidding in traction movement mode

The parameters on Fig.1 are as follows:

$P_{f_1}$ - rolling resistance of front driven wheels;

$\delta_1$ - side slip angle of the front axle;

$R_{\delta_1}$ - the total lateral response to the front axle;

$P_{W_{X1}}$ and $P_{W_{Y1}}$ - are longitudinal and lateral components of the air resistance force;

ψ - mobile vehicle course angle;

*a* and *b* - the distance from the front and rear axle, respectively, to the projection of the mobile vehicle center of mass on a horizontal plane;

*L* - mobile vehicle longitudinal wheelbase.

Vector $\overline{R}_B$ of overall reaction in the plane of the road on the rear axle is collinear to the relative speed of the rear wheels slip, $\overline{V}_B$ but oppositely directed [19] (see Fig. 1).

The vector $\overline{V}_B$ is the sum of two vectors [12]

$$\overline{V}_B = \overline{V}_{slip} + \overline{V}_{B/O}, \tag{1}$$

where $\overline{V}_{slip}$ - drive wheels slipping speed.

$$V_{slip} = V_{circ} - V_{X1} = S_x \cdot V_{circ}; \tag{2}$$

$\overline{V}_{circ}$ - the peripheral speed of the wheel;

$\overline{V}_{B/O}$ - linear velocity of the point *B* with respect to the instantaneous rotation center *O* (see Fig. 1);

$S_x$ - the relative slippage of the drive wheels,

$$S_x = \frac{V_{slip}}{V_{circ}} = \frac{V_{circ} - V_{X1}}{V_{circ}} = 1 - \frac{V_{X1}}{V_{circ}}. \tag{3}$$

The angle between the $\overline{R}_B$ vector and $CY_1$ axis [12]

$$tg\gamma_B = \frac{ctg\Theta_B}{1 - S_x}. \tag{4}$$

The system of differential equations describing the motion of a mobile vehicle, in this case, is the following

$$\begin{cases} m_a \dfrac{d^2 x_1}{dt^2} = -P_{f_1} + R_{\delta_1} \cdot \sin\delta_1 + \\ +R_B \cdot \sin\gamma_B - P_{W_{X1}}; & (5) \\ m_a \dfrac{d^2 y_1}{dt^2} = -R_{\delta_1} \cdot \cos\delta_1 - R_B \cdot \cos\gamma_B + P_{W_{Y1}}; & (6) \\ I_{zc} \dfrac{d\omega_z}{dt} = R_{\delta_1} \cdot a \cdot \cos\delta_1 - R_B \cdot b \cdot \cos\gamma_B, & (7) \end{cases}$$

where $I_{zc}$ - moment of inertia of the mobile vehicle relative to the vertical axis passing through the center of mass [5];

$\omega_z$ - angular velocity of the mobile vehicle in the plane of the road;

$m_a$ - weight of the mobile vehicle;

$\dfrac{d^2 x_1}{dt^2}$, $\dfrac{d^2 y_1}{dt^2}$ - linear acceleration of the mobile vehicle, respectively, in the longitudinal and transverse planes.

Acceleration of the mobile vehicle along the axes $OX_1$ and $OY_1$ are defined as

$$\begin{cases} \dfrac{d^2 x_1}{dt^2} = a_c^n \cdot \sin\Theta_C + a_c^k \cdot \cos\Theta_C; & (8) \\ \dfrac{d^2 y_1}{dt^2} = a_c^k \cdot \sin\Theta_C - a_c^n \cdot \cos\Theta_C, & (9) \end{cases}$$

where $a_C^n, a_C^k$ - normal and tangential component of acceleration of the mobile vehicle center of mass;

$\Theta_C$ - the angle between the vector $\overline{a_c^n}$ and axis $OY_1$

$$tg\Theta_C = tg\delta_1 + \frac{a \cdot \omega_z}{V_{X_1}}. \qquad (10)$$

The normal and tangential component of acceleration of the mobile vehicle center of mass

$$a_c^n = \frac{\omega_z \cdot V_{X_1}}{\cos\Theta_C}; \qquad (11)$$

$$a_c^k = \frac{d\omega_z}{dt} \cdot \frac{V_{X_1}}{\omega_z \cdot \cos\Theta_C}. \qquad (12)$$

After substituting equations (11) and (12) into the system of equations (8) and (9) we get by (10)

$$\begin{cases} \dfrac{d^2 x_1}{dt^2} = a \cdot \omega_z^2 + \omega_z \cdot V_{X_1} \cdot tg\delta_1 + \dfrac{d\omega_z}{dt} \dfrac{V_{X_1}}{\omega_z}; & (13) \\ \dfrac{d^2 y_1}{dt^2} = a \dfrac{d\omega_z}{dt} + \dfrac{d\omega_z}{dt} \dfrac{V_{X_1}}{\omega_z} \cdot tg\delta_1 - \omega_z \cdot V_{X_1}. & (14) \end{cases}$$

Using the expressions (13) and (14) the system of equations (5) - (7) takes the form

$$\begin{cases} m_a \cdot \left( a \cdot \omega_z^2 + \omega_z \cdot V_{X_1} \cdot tg\delta_1 + \dfrac{d\omega_z}{dt} \cdot \dfrac{V_{X_1}}{\omega_z} \right) = \\ = -P_{f_1} + R_{\delta_1} \sin\delta_1 + R_B \sin\gamma_B - P_{W_{X1}}; & (15) \\ m_a \cdot \left( a \cdot \dfrac{d\omega_z}{dt} + \dfrac{d\omega_z}{dt} \cdot \dfrac{V_{X_1}}{\omega_z} \cdot tg\delta_1 - \omega_z \cdot V_{X_1} \right) = \\ = -R_{\delta_1} \cos\delta_1 - R_B \cos\gamma_B + P_{W_{Y1}}; & (16) \\ I_{zc} \cdot \dfrac{d\omega_z}{dt} = R_{\delta_1} \cdot a \cdot \cos\delta_1 - R_B \cdot b \cdot \cos\gamma_B. & (17) \end{cases}$$

Solving equations (16) and (17), we define the angular acceleration of the mobile vehicle

$$\frac{d\omega_z}{dt} = \frac{a \cdot \omega_z \cdot V_{X_1} - L \cdot \dfrac{R_B}{m_a} \cdot \cos\gamma_B + a \cdot \dfrac{P_{W_{Y1}}}{m_a}}{a^2 + i_z^2 + a \cdot \dfrac{V_{X_1}}{\omega_z} \cdot tg\delta_1}, \qquad (18)$$

where $i_z$ - radius of inertia of the mobile vehicle relative to the vertical axis,

$$i_z = \sqrt{\frac{I_{zc}}{m_a}}. \qquad (19)$$

Total vertical reaction in the plane of the road wheels on the rear axle during traction movement mode

$$R_{z_2} = m_a \cdot g \cdot \frac{a}{L} + P_w \cdot \frac{h_w - r_\partial}{L} + P_j \cdot \frac{h - r_\partial}{L}, \qquad (20)$$

where $P_j$ - the force of inertia;

$h_w$ - the height of the center of aerodynamic pressure.

Assuming the turning point of the air resistance force is zero (metacentre position coincides with the position of the center of mass), let $h_w \approx h$.

Total tangent reaction of the road in response to the rear axle can be determined from the expression

$$R_B = \varphi \cdot R_{z_2}, \tag{21}$$

where φ - coefficient of grip.

Substituting the relationship (13) and (20) into (21)

$$R_B = m_a \cdot \varphi \cdot \left[ g\frac{a}{L} + \left( a\omega_z^2 + \omega_z V_{X_1} tg\delta_1 + \frac{d\omega_z}{dt} \cdot \frac{V_{X_1}}{\omega_z} + \frac{k \cdot F}{m_a} \cdot V_{X_1}^2 \right) \cdot \frac{h - r_\partial}{L} \right]. \tag{22}$$

After substitution of (22) in (18) we finally obtain

$$\frac{d\omega_z}{dt} = \frac{a \cdot \left( \omega_z \cdot V_{X_1} + \frac{P_{W_{Y1}}}{m_a} \right)}{a^2 + i_z^2 + \frac{V_{X_1}}{\omega_z} \cdot \left( a \cdot tg\delta_1 + \Omega \right)} - \frac{\frac{ag\Omega}{h - r_\partial} - \left( a\omega_z^2 + \omega_z V_{X_1} tg\delta_1 + \frac{kF}{m_a} V_{X_1}^2 \right)\Omega}{a^2 + i_z^2 + \frac{V_{X_1}}{\omega_z} \cdot \left( a \cdot tg\delta_1 + \Omega \right)}, \tag{23}$$

where $\Omega = \varphi \cdot \cos\gamma_B \cdot \left( h - r_\partial \right)$.

Air resistance force $P_W$, as well as other components of the resultant aerodynamic forces and moments increase as a square of the velocity of the mobile vehicle [4]

$$P_{W_{X1}} = c_X \cdot \frac{\rho}{2} \cdot F \cdot V^2; \tag{24}$$

$$P_{W_{Y1}} = c_Y \cdot \frac{\rho}{2} \cdot F \cdot V^2, \tag{25}$$

where $c_X$ and $c_Y$ - drag coefficient in the longitudinal and transverse planes;

$F$ - the projected area of the mobile vehicle on a plane perpendicular to the longitudinal axis;

ρ - the density of the ambient mobile vehicle air.

The skid damping is possible with $\frac{d\omega_Z}{dt} \leq 0$.

Fig. 2 shows the angular acceleration $\varepsilon_Z = \frac{d\omega_Z}{dt}$ of conventional rear-drive mobile vehicle in the plane of the road from some of the parameters of DMVRE system.

Parameters and conditions of the mobile vehicle are the following: $a$ = 1,2 m; $L$ = 2,5 m; $kF$ = 0,58 Hs$^2$/m$^2$; $h$ = 0,5 m; $r_\partial$ = 0,29 m; ρ = 1,22 kg/m$^3$; $i_z$ = 1,3 m; $m$ = 1500 kg; g = 9,81 m/s$^2$.

Analysis of Fig. 2 shows that the growth of the initial skidding speed of mobile vehicle $\varepsilon_Z$ increases, and with increasing of slip angle $\varepsilon_Z$ is reduced. Reduction of the grip rate, increasing of angular speed rate and relative slippage of the drive wheels increase $\varepsilon_Z$.

During rear-drive mobile vehicle skidding that moved at speeds below $V_{X_{stab}}$ (see Fig. 2a, 2b, 2d), at the initial introduction, a negative acceleration, stabilizing angle on the mobile vehicle is arises.

This fact indicates the presence of self-sustainability ("immunity") of mobile vehicle against skidding. If the skidding of mobile vehicle, moving at speeds above $V_{X_{stab}}$, their own stability is not enough to stabilize, so the driver must intervene or the stability system needs to be triggered.

For the simulated mobile vehicle (Fig. 2a) $V_{X_{stab}}$ = 0...24 km/h with $\omega_Z$ = 0,1 s$^{-1}$, depending on the tire slip angles.

So, using the relation (23) it's possible to determine the limiting values of the DMVRE parameters on condition of mobile vehicle stability.

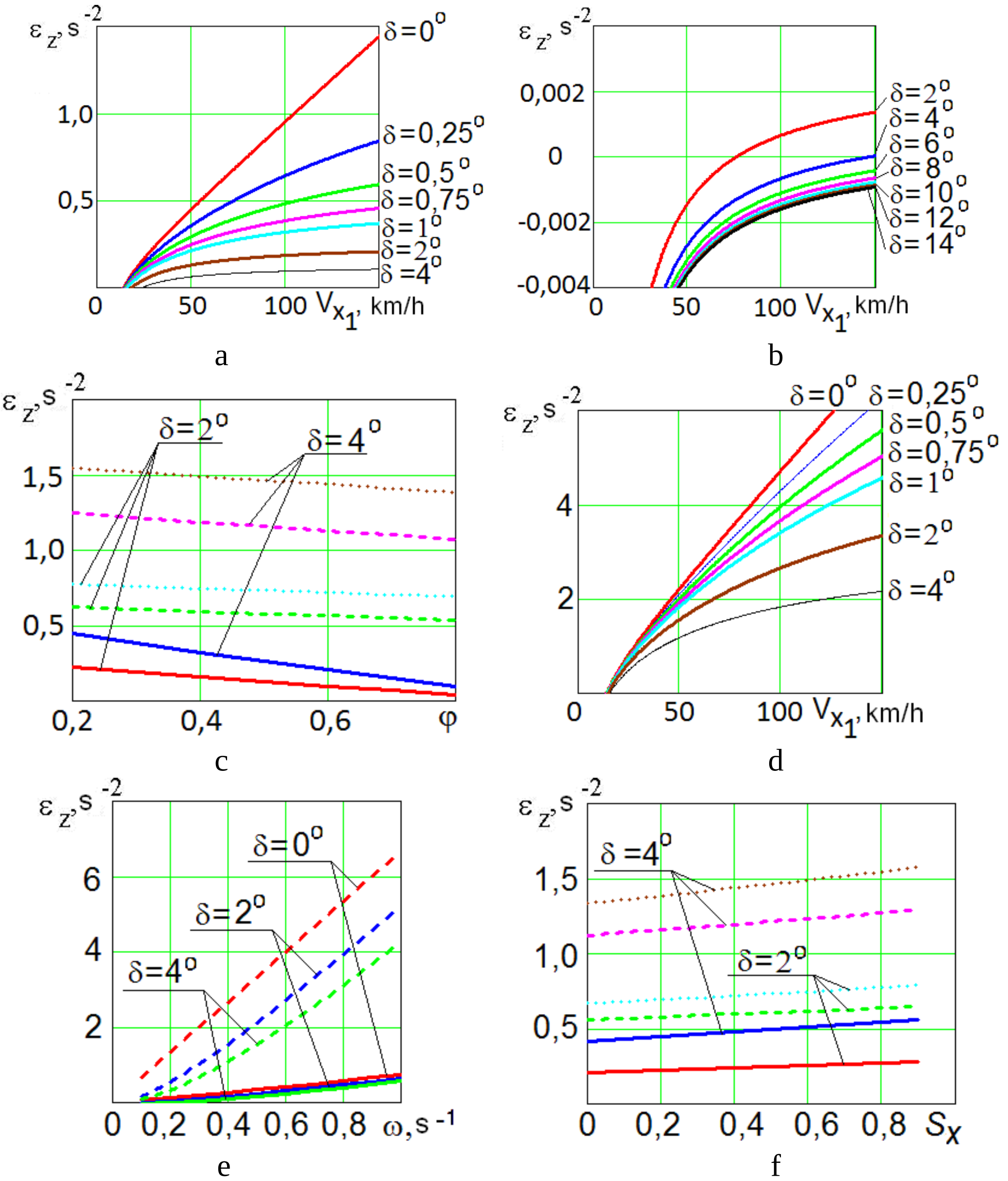

Fig. 2 - Dependence of the rear-wheel mobile vehicle angular acceleration of in the plane of the road from some of the parameters of DMVRE system: a - dependence $\varepsilon_z(V_{X_1})$ at = 0.1 s$^{-1}$; b - dependence $\varepsilon_z(V_{X_1})$ at $\omega_z$ = 0.01 s$^{-1}$; c - dependence $\varepsilon_z(\varphi)$ at $\omega_z$ = 0.2 s$^{-1}$, d - dependence $\varepsilon_z(V_{X_1})$ at $\omega_z$ = 0.5 s$^{-1}$; e - dependence $\varepsilon_z(\omega_z)$ at various angles of diversion, f - dependence $\varepsilon_z(S_X)$ with different slip angles; Fig. 2c, 2d and 2e have the following designations: ___ $V_{X_1}$ = 5 m/s --- $V_{X_1}$ = 20 m/s, ▪ ▪ ▪ ▪ $V_{X_1}$ = 35 m/s

## CONCLUSIONS

1. With the growing of initial skidding speed of mobile vehicles their angular acceleration in the plane of the road $\varepsilon_z$ increases, and with slip angle increasing $\varepsilon_z$ is reduced. Reduction of the grip rate, increasing of angular speed rate and relative slippage of the drive wheels increase $\varepsilon_z$.

2. The dependences obtained allow to determine the velocity intervals $V_{X_{stab}}$. which can achieve the negative acceleration at the initial skid moment, stabilizing the vehicle trajectory. As mobile vehicle is drifting at speeds above $V_{X_{stab}}$, its own stability is not enough to stabilize, so the driver must intervene or the stability system needs to be triggered. For the studied mobile vehicle $V_{X_{stab}}$ = 0...24 km/h with $\omega_z$ = 0,1 s$^{-1}$, depending on the tire slip angles.

## МОДЕЛЮВАННЯ ПРОЦЕСУ ЗАНОСУ МОБІЛЬНИХ МАШИН В ТЯГОВОМУ РЕЖИМІ РУХУ

**Анотація.** В роботі запропоновано математичну модель, що пов’язує між собою параметри системи «водій - мобільна машина - дорожнє середовище» та дозволяє моделювати процес заносу мобільних машин в тяговому режимі руху. Використання нелінійних моделей процесу заносу дозволяє розробити ефективні алгоритми роботи систем динамічної стабілізації мобільних машин.